\documentclass[prb,twocolumn,amsmath,amssymb,superscriptaddress]{revtex4-2}
\usepackage[colorlinks = true,
            linkcolor = red,
            urlcolor  = red,
            citecolor = blue,
            anchorcolor = blue]{hyperref}
\usepackage{epsfig}
\usepackage{amsmath,amssymb}
\usepackage{physics}
\usepackage{graphicx}
\usepackage[dvipsnames,usenames]{xcolor}
\usepackage[normalem]{ulem}
\usepackage{todonotes}
\usepackage{soul} 
\usepackage{orcidlink}
\usepackage{glossaries}
\usepackage{hyperref}
\newacronym{DQMC}{DQMC}{determinant quantum Monte Carlo}
\newacronym{QMC}{QMC}{Quantum Monte Carlo}
\newacronym{HMC}{HMC}{hydrid Monte Carlo}
\newacronym{eph}{$e$-ph}{electron-phonon}
\newacronym{ee}{$e$-$e$}{electron-electron}
\newacronym{AFM}{AFM}{antiferromagnetic}
\newacronym{BOW}{BOW}{bond-order-wave}
\newacronym{CDW}{CDW}{charge-density-wave}
\newacronym{SSH}{SSH}{Su-Schrieffer-Heeger}
\newacronym{OSSH}{oSSH}{``optical" SSH}
\newacronym{BSSH}{bSSH}{``bond" SSH}
\newacronym{2D}{2D}{two-dimensional}
\newacronym{QCP}{QCP}{quantum critical point}

\begin{document}

\title{Antiferromagnetic and bond-order-wave phases in the half-filled two-dimensional optical Su-Schrieffer-Heeger-Hubbard model}

\author{Andy~{Tanjaroon~Ly}\orcidlink{0000-0003-3162-8581}}
\affiliation{Department of Physics and Astronomy, The University of Tennessee, Knoxville, TN 37996, USA}
\affiliation{Institute for Advanced Materials and Manufacturing, The University of Tennessee, Knoxville, TN 37996, USA\looseness=-1} 

\author{Benjamin~Cohen-Stead\orcidlink{0000-0002-7915-6280}}
\affiliation{Department of Physics and Astronomy, The University of Tennessee, Knoxville, TN 37996, USA}
\affiliation{Institute for Advanced Materials and Manufacturing, The University of Tennessee, Knoxville, TN 37996, USA\looseness=-1} 

\author{Steven Johnston\orcidlink{0000-0002-2343-0113}}
\affiliation{Department of Physics and Astronomy, The University of Tennessee, Knoxville, TN 37996, USA}
\affiliation{Institute for Advanced Materials and Manufacturing, The University of Tennessee, Knoxville, TN 37996, USA\looseness=-1} 

\begin{abstract}
Electron-phonon ($e$-ph) interactions arise in many strongly correlated quantum materials from the modulation of the nearest-neighbor hopping integrals, as in the celebrated \gls*{SSH} model. Nevertheless, relatively few non-perturbative studies of correlated SSH models have been conducted in dimensions greater than one, and those that have been done have primarily focused on bond models, where generalized displacements independently modulate each hopping integral. We conducted a sign-problem free determinant quantum Monte Carlo study of the optical SSH-Hubbard model on a two-dimensional square lattice, where site-centered phonon modes simultaneously modulate pairs of nearest-neighbor hopping integrals. We report the model's low-temperature phase diagram in the challenging adiabatic regime ($\Omega/E_\mathrm{F} \sim 1/8$). It exhibits insulating antiferromagnetic Mott and bond-order-wave (BOW) phases with a narrow region of coexistence between them. We also find that a critical $e$-ph coupling is required to stabilize the BOW phase in the small $U$ limit. Lastly, in stark contrast to recent findings for the model's bond variant, we find no evidence for a long-range antiferromagnetism in the pure $(U/t=0)$ optical \gls*{SSH} model. 
\end{abstract} 

\maketitle



\section{Introduction}\label{sec:introduction}

\noindent Strongly correlated quantum materials are often governed by the interplay between two or more degrees of freedom, which compete or cooperate with one another to produce novel states of matter and complex phase diagrams~\cite{Dagotto2005complexity, Morosan2012strongly, Davis2013concepts, Fradkin2015colloquium}. For example, many interesting phases can be derived from combining \gls*{eph} and \gls*{ee} interactions. A canonical model for this physics is the half-filled Hubbard-Holstein model, which hosts charge-density-wave, \gls*{AFM}, and intermediate metallic phases, with the latter depending on the dimensionality of the system~\cite{Scalettar1989competition, Koller2004HubbardHolstein, Werner2007HubbardHolstein, Johnston2013HubbardHolstein, Nowadnick2015renormalization, Karakuzu2017superconductivity, Wang2020HHPhases, Costa2020HHPhases}. There is also evidence suggesting that \gls*{eph} interactions can enhance $d$-wave pairing~\cite{Zeyher1996renormalization, Huang2003electronphonon, Nowadnick2015renormalization} and affect spin and charge correlations~\cite{Karakuzu2022stripe, Liu2024charge} in doped Hubbard-Holstein models. 

Recently, interest in \gls*{eph} interactions has shifted toward \gls*{SSH} models, where the atomic motion modulates the electronic hopping integrals~\cite{Barisic1970tightbinding, Su1979solitons}. While different variants of the \gls*{SSH} model have been introduced over the years, a widely studied one is the so-called \gls*{BSSH} model~\cite{Sengupta2003Peierls}, which places independent harmonic oscillators on the bonds between atomic sites. \gls*{QMC} studies of the \gls*{BSSH} model have yielded interesting results regarding superconductivity~\cite{Zhang2023, cai2023hightemperature} and magnetism, including the presence of antiferromagnetism absent a Coulomb interaction~\cite{Cai2021antiferromagnetism, Gotz2022Valence}. Subsequent work reported on the robustness of this anomalous antiferromagnetism and its competition with a \gls*{BOW} phase, both with and without a Hubbard interaction~\cite{Cai2022robustness, Feng2022phase, Xing2023attractive}. A lesser studied variant is the \gls*{OSSH} model~\cite{Capone1997Optical}, which instead places dispersionless Einstein phonons on the atomic sites that simultaneously modulate all of the surrounding nearest-neighbor hopping integrals. Recent \gls*{DQMC} studies have shown that the two models produce quite different physical results~\cite{TanjaroonLy2023comparative, MalkarugeCosta2023comparative}, which motivates studies into correlated \gls*{OSSH} models. This class of interactions has also been implicated in a wide range of quantum materials~\cite{Zhao2000, Devereaux2004, Johnston2010, Cohen-Stead2022, Zhang2014, Johnston2014}. 

This article presents a detailed study of the low temperature phases of the \gls*{OSSH}-Hubbard model on a square lattice using numerically exact \gls*{DQMC}. Both the \gls*{OSSH} and \gls*{BSSH} couplings preserve particle-hole symmetry at half-filling such that the simulations remain sign-problem-free, unlike in the Hubbard-Holstein model~\cite{Johnston2013HubbardHolstein}. Focusing on the adiabatic regime ($\Omega/E_{\mathrm{F}}\sim 1/8$) relevant to most materials, we obtain a low-temperature phase diagram with {$\boldsymbol{Q} = (\pi/a,\pi/a)$ \gls*{AFM} and \gls*{BOW} phases, depending on the relative strengths of the \gls*{ee} and \gls*{eph} interactions. We also find evidence for a narrow region of coexistence between these phases. At $T = 0$, the \gls*{AFM} and \gls*{BOW} phases are insulating; however, we find a small region of metallically for small values of $(\lambda, U)$ at finite temperatures but no indications of enhanced superconductivity in this region. Importantly, we also find no evidence for an \gls*{AFM} phase in the pure \gls*{OSSH} model ($U=0$), in contrast to the \gls*{BSSH} model~\cite{Cai2021antiferromagnetism, Gotz2022Valence}. Instead, we find only a weak enhancement of short-range \gls*{AFM} correlations in this region of parameter space. Our results highlight crucial differences between the two variants of the \gls*{SSH} model and how it competes with Mott correlations. \\



\section{Model \& Methods}\label{sec:methods}
We study the \gls*{OSSH}-Hubbard model defined on two-dimensional $N = L\times L$ square lattices with periodic boundary conditions. Unlike the more commonly studied bond model~\cite{Sengupta2003Peierls}, our model places the phonons on the sites~\cite{Capone1997Optical} such that each displacement simultaneously modulates two neighboring bonds rather than one. The model's Hamiltonian is
\begin{equation}\label{eq:H}
    \hat{H} = \hat{H}_e + \hat{H}_\text{ph} + \hat{H}_U + 
    \hat{H}_{e-\text{ph}},
\end{equation}
where 
\begin{equation}\label{eq:H0}
\hat{H}_e=-t\sum_{\boldsymbol{i}, \nu,\sigma} (\hat{c}^\dagger_{{\boldsymbol{i}}+{\boldsymbol a}_\nu,\sigma}\hat{c}^{\phantom\dagger}_{{\boldsymbol{i}},\sigma} + \textrm{H.c.})- \mu\sum_{{\boldsymbol{i}},\sigma} \hat{n}_{{\boldsymbol{i}},\sigma}
\end{equation}
are the non-interacting electron terms, 
\begin{equation}
\hat{H}_\text{ph} = \sum_{\boldsymbol{i},\nu}\bigg( \frac{1}{2M}\hat{P}_{\boldsymbol{i},\nu}^2+\frac{M\Omega^2}{2} \hat{X}_{\boldsymbol{i},\nu}^2 \bigg)
\end{equation}
are the non-interacting phonon terms, 
\begin{equation}
\hat{H}_U = U\sum_{\boldsymbol{i}}\hat{n}_{\boldsymbol{i},\uparrow}\hat{n}_{\boldsymbol{i},\downarrow}
\end{equation}
is the Hubbard interaction, and
\begin{equation}\label{eq:Heph}
\hat{H}_{e-\text{ph}}=\alpha\sum_{\boldsymbol{i},\nu,\sigma}(\hat{X}_{\boldsymbol{i}+\boldsymbol{a}_\nu,\nu}-\hat{X}_{\boldsymbol{i},\nu}) (\hat{c}^\dagger_{{\boldsymbol{i}}+{\boldsymbol{a}}_\nu,\sigma}\hat{c}^{\phantom\dagger}_{{\boldsymbol{i}},\sigma} + \textrm{H.c.}).
\end{equation}
is the \gls*{eph} interaction. In Eqs.~\eqref{eq:H0}-\eqref{eq:Heph}, $\hat{c}^\dagger_{{\boldsymbol{i}},\sigma}$ ($\hat{c}^{\phantom\dagger}_{{\boldsymbol{i}},\sigma}$) creates (annihilates) a spin-$\sigma$ ($=\uparrow,\downarrow$) electron at lattice site ${\boldsymbol{i}}$,  $\hat{n}_{\boldsymbol{i},\sigma} = \hat{c}^\dagger_{\boldsymbol{i},\sigma}\hat{c}^{\phantom\dagger}_{\boldsymbol{i},\sigma}$ is the number operator, and $\text{H.c.}$ denotes the Hermitian conjugate. 
The index $\nu$ runs over the $\hat{x}$ and $\hat{y}$ spatial dimensions, and $\boldsymbol{a}_x=(a,0)$ and $\boldsymbol{a}_y = (0,a)$ are the primitive lattice vectors of the square lattice. $t$ is the nearest neighbor hopping integral, $\mu$ is the chemical potential, $U$ is the Hubbard repulsion, $M$ is the ion mass, and $\hbar\Omega$ is the phonon energy. Lastly, $\hat{X}_{\boldsymbol{i}, \nu}$ ($\hat{P}_{\boldsymbol{i}, \nu}$) is the position (momentum) operator for the atom at site ${\boldsymbol{i}}$ along the $\nu=\hat{x},\hat{y}$ directions, and $\alpha$ controls the \gls*{eph} coupling strength. 

We solve Eq~\eqref{eq:H} using numerically exact \gls*{DQMC}, as implemented in the \texttt{SmoQyDQMC.jl} package~\cite{White1989, SmoQyDQMC1, SmoQyDQMC2}. Throughout this work, we focus on half-filling $\langle n \rangle$ = 1, and take $t=M=a=\hbar=1$ to set our units of measurement. We further fix $\Omega = t/2$ unless otherwise stated and adopt a Brillouin zone-averaged definition of the dimensionless coupling 
\begin{align}
    \lambda&= \frac{2}{W N^2}\sum_{\boldsymbol{k},\boldsymbol{q}, \nu} \frac{|g_\nu(\boldsymbol{k},\boldsymbol{q})|^2}{\Omega(\boldsymbol{q})}
    = \frac{8\alpha^2}{M\Omega^2 W}, 
    \label{eq:lambda}
\end{align}
where $g_\nu(\boldsymbol{k},\boldsymbol{q})$ is the \gls*{eph} matrix element for phonon branch $\nu$, $W = 8t$ is the non-interacting bandwidth, and $E_\mathrm{F} = 4t$ is the Fermi energy. We refer the reader to Refs.~\cite{TanjaroonLy2023comparative, MalkarugeCosta2023comparative} for further discussion on defining $\lambda$ and the explicit form of $g(\boldsymbol{k},\boldsymbol{q})$. 

We assess the model's ordering tendencies by 
measuring the susceptibilities 
\begin{equation}
    \chi_{\gamma}(\boldsymbol{q})=\int_{0}^{\beta} S_\gamma(\boldsymbol{q},\tau) \mathrm{d}\tau, 
\end{equation}
where $\gamma$ indexes the spin ($\text{s}$), bond ($\text{b}$), or pairing ($\text{p}$) correlations 
and 
\begin{equation}
S_{\gamma}(\boldsymbol{q},\tau)=\frac{1}{N}\sum_{{\boldsymbol{i}},{\bf j}}e^{-\mathrm{i}\boldsymbol{q}\cdot(\boldsymbol{R}_{\boldsymbol{i}}-\boldsymbol{R}_{\bf j})}\left\langle \hat{O}^{\phantom\dagger}_{\gamma,{\boldsymbol{i}}}(\tau)\hat{O}_{\gamma,{\bf j}}^{\dagger}(0)\right\rangle 
\end{equation}
is the corresponding structure factor. The relevant operators are  
\begin{equation}
\hat{O}_{\text{s},\boldsymbol{i}} = \frac{1}{2}\left(\hat{n}_{\boldsymbol{i},\uparrow}-\hat{n}_{\boldsymbol{i},\downarrow}\right), 
\end{equation}
for \gls*{AFM} correlations, 
\begin{equation}
\hat{O}_{\mathrm{b},\boldsymbol{i}}  =  \frac{1}{2}\sum_{\sigma}(\hat{c}^\dagger_{{\boldsymbol{i}},\sigma}\hat{c}^{\phantom\dagger}_{{\boldsymbol{i}}+\boldsymbol{a}_\nu,\sigma} + \textrm{H.c.}), 
\end{equation}
for bond correlations and 
\begin{subequations}
\begin{align}
    \hat{O}_{\text{p},\boldsymbol{i}}&= \hat{c}_{{\boldsymbol{i}},\uparrow}\hat{c}_{{\boldsymbol{i}},\downarrow}, \\
    \hat{O}_{\text{p},\boldsymbol{i}}^{s^*} &= \frac{1}{2}\sum_{\nu}(\hat{c}_{{\boldsymbol{i}},\uparrow}\hat{c}_{{\boldsymbol{i}}+\boldsymbol{a}_{\nu},\downarrow} + \hat{c}_{{\boldsymbol{i}},\uparrow}\hat{c}_{{\boldsymbol{i}}-\boldsymbol{a}_{\nu},\downarrow}),~\text{and} \\
    \hat{O}_{\text{p},\boldsymbol{i}}^{d}&= \frac{1}{2}\sum_{\nu}\mathcal{P}_\nu(\hat{c}_{{\boldsymbol{i}},\uparrow}\hat{c}_{{\boldsymbol{i}}+\boldsymbol{a}_{\nu},\downarrow} + \hat{c}_{{\boldsymbol{i}},\uparrow}\hat{c}_{{\boldsymbol{i}}-\boldsymbol{a}_{\nu},\downarrow})
\end{align}
\end{subequations}
for the superconducting correlations with $s$-, extended $s$-, and $d$-wave symmetries, respectively. Here, $\mathcal{P}_\nu = (\delta_{x,\nu}-\delta_{y,\nu})$ is a phase factor for the $d$-wave symmetry.

To assess critical behavior, we compute the correlation ratio from the equal-time structure factors
\begin{equation*}
    R_{\gamma}(L) = 1 - \frac{S_{\gamma}(\boldsymbol{Q}-\delta\boldsymbol{q}_L,0)}{S_{\gamma}(\boldsymbol{Q},0)},
\end{equation*}
where $L$ denotes the cluster size, $|\delta\boldsymbol{q}_L|=\frac{2\pi}{La}$ and $\boldsymbol{Q}=(\pi/a,\pi/a)$ is the ordering wave for both the \gls*{AFM} and \gls*{BOW} correlations. The critical point for the transition can then be determined from the crossing point of the $R_\gamma(L)$ curves for different lattice sizes while fixing $\beta= L$ to access low-temperature (ground state) properties~\cite{KaulPRL2015triangular, SatoPRL2018kondoqmc, LiuPRB2018qcp, DarmawanPRB2018stripevmctensor}. We also performed measurements of other standard quantities, such as the double occupancy and average electron kinetic energy 
\begin{equation}
    \langle K_{\nu} \rangle = -\left\langle 
    t(\hat{X}_{\boldsymbol{i}+\boldsymbol{a}_\nu,\nu}-\hat{X}_{\boldsymbol{i},\nu})
    \hat{c}_{\boldsymbol{i} + \boldsymbol{a}_\nu, \sigma}^\dagger \hat{c}_{\boldsymbol{i},\sigma}^{\phantom{\dagger}} + \text{H.c.} \right\rangle. 
\end{equation}

When computing the local observables, we bias the system for the $x$-bonds to ensure that we select the same broken symmetry state. \\
\\

\begin{figure}[t]
    \centering
    \includegraphics[width=1.0\columnwidth]{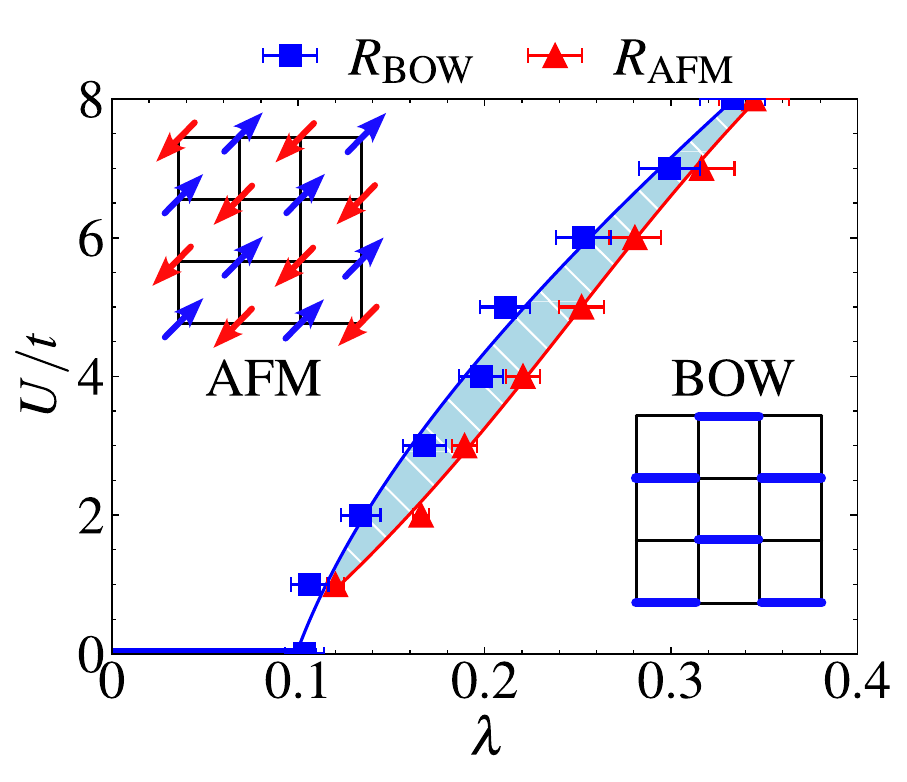}
    \caption{Phase diagram of the optical SSH-Hubbard model at half-filling $\langle n\rangle = 1$ with a fixed phonon frequency $\Omega /t = 0.5$. The red triangles and blue squares denote the transition points obtained from \gls*{AFM} and \gls*{BOW} correlation ratios, respectively, with solid lines providing a guide for the eye. The blue hatched area represents the region of \gls*{AFM}/\gls*{BOW} coexistence. The thick blue line from $\lambda = 0$ to $\lambda_\mathrm{c}$ along the $U=0$ line denotes a region where the pure optical SSH model has no long-range \gls*{AFM} or \gls*{BOW} order.}
    \label{fig:figure1_phase_diagram}
\end{figure}

\section{Results}\label{sec::results}

Figure~\ref{fig:figure1_phase_diagram} plots the low-temperature phase diagram of the half-filled \gls*{OSSH}-Hubbard model in the $\lambda$-$U$ plane. For small \gls*{eph} coupling, the system is dominated by \gls*{AFM} Mott/Slater  correlations depending on the value of $U/t$. Increasing the \gls*{eph} coupling for fixed $U/t$ drives the system into a non-magnetic insulating \gls*{BOW} phase, with a narrow intermediate regime of \gls*{AFM}/\gls*{BOW} coexistence, as indicated by the hatched shading. 
Additionally, in the $U/t = 0$ limit, the \gls*{BOW} ground state disappears below a critical coupling $\lambda_c^\text{BOW} = 0.103 \pm 0.010$. Below this coupling, we observe a weak enhancement of the short-range \gls*{AFM} correlations in the pure \gls*{OSSH} model but no long-range \gls*{AFM} order. This result is in direct contrast with the \gls*{BSSH} model, where long-range magnetic order was reported~\cite{Cai2021antiferromagnetism, Gotz2022Valence}. We can understand this result by considering the difference in the effective $e$-$e$ interactions that arise in the anti-adiabatic limits of the two models, see Supplementary Note 1~\cite{supplement}. We also estimate that for $\lambda < \lambda_c^\text{BOW}$ and $U/t\lesssim 2$, the system is an \gls*{AFM} metal at elevated temperatures ($\beta \ge 16/t$, see Supplementary Note 2) before transitioning to an insulator at higher $U/t$ or lower temperatures.  
In the following sections, we will discuss the various measurements that lead to this phase diagram.\\

The phase boundaries in Fig.~\ref{fig:figure1_phase_diagram} were obtained from the crossing point of the \gls*{BOW} and \gls*{AFM} correlation ratios, as illustrated in Fig.~\ref{fig:figure2_scaling}. Here we show the correlation ratio for \gls*{BOW} and \gls*{AFM} order in Figs~\ref{fig:figure2_scaling}a and \ref{fig:figure2_scaling}b, respectively, for a representative value of $U/t = 3$. This analysis is based on a finite-size scaling analysis~\cite{SatoPRL2018kondoqmc, LiuPRB2018qcp, DarmawanPRB2018stripevmctensor}, where the \gls*{QCP} $\lambda_c$ corresponds to the crossing point of the $R_{\gamma}(L)$ curves for different cluster sizes $L$ while fixing $\beta t = L$. This approach assumes that the dynamical critical exponent for the \gls*{AFM} and \gls*{BOW} transitions is $z = 1$~\cite{Costa2020HHPhases, Goetz2024phases}. In this case, we obtain $\lambda_c^{\mathrm{BOW}}=0.16 \pm 0.01$ (see Fig.~\ref{fig:figure2_scaling}a) and $\lambda_c^{\mathrm{AFM}} =  0.189\pm 0.007$ (see Fig.~\ref{fig:figure2_scaling}b). The difference in these values suggests a narrow region of coexistence between the two states, where \gls*{BOW} order has begun to set in while \gls*{AFM} order has not yet disappeared. This result is distinct from that of the \gls*{BSSH} and \gls*{BSSH}-Hubbard models, in which a direct transition from a \gls*{BOW} to \gls*{AFM} ground state occurs, with no reported region of coexistence~\cite{Cai2021antiferromagnetism, Gotz2022Valence, Cai2022robustness, Feng2022phase}. The remaining points in Fig.~\ref{fig:figure1_phase_diagram} were obtained from a similar analysis, as shown in Supplementary Note 3~\cite{supplement}. 

\begin{figure}[t]
    \centering
    \includegraphics[width=\columnwidth]{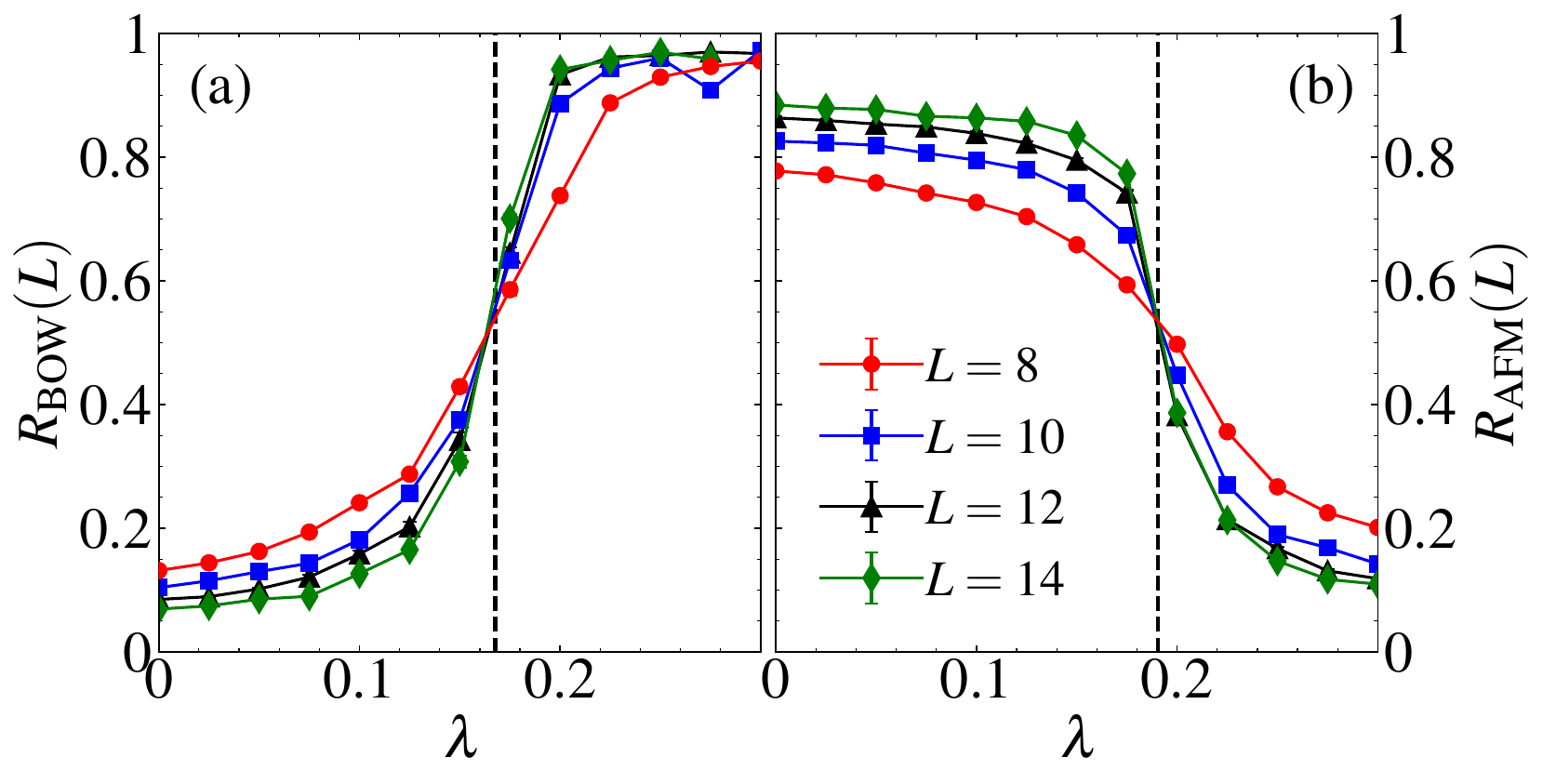}
    \caption{Correlation ratios for the (a) bond-ordered-wave and (b) antiferromagnetic phases as a function of $\lambda$ for fixed $U/t = 3$. The location of the quantum critical point $\lambda_c$ is obtained from the crossing of the correlation ratio for different lattice sizes $L$ and is indicated by the dashed lines. All calculations were performed for fixed phonon frequency $\Omega /t = 0.5$ and at an inverse temperature $\beta t = L$. The definition of the correlation ratio is provided in the Methods section. 
    \label{fig:figure2_scaling}}
\end{figure}

Another key difference between our phase diagram and results obtained for the \gls*{BSSH} model pertains to the behavior along the $U/t = 0$ line. In the pure \gls*{BSSH} model, \gls*{AFM} order is stabilized without a Hubbard interaction~\cite{Cai2021antiferromagnetism}. Here, however, we observe no such order along the $U/t = 0$ line but instead only a weak enhancement of short-range \gls*{AFM} correlations, see Supplementary Note 4. This behavior is illustrated in Figs.~\ref{fig:figure3_corr_ratio_U0}a and \ref{fig:figure3_corr_ratio_U0}b, which plot the \gls*{BOW} and \gls*{AFM} correlation ratios in this limit, respectively. While we obtain a finite critical coupling for the \gls*{BOW} phase from the corresponding correlation ratio crossing, we observe no similar crossing in \gls*{AFM} correlation ratio. Taking  a larger $\Omega/t = 2$ produces similar results, as shown in Figs.~\ref{fig:figure3_corr_ratio_U0}c and \ref{fig:figure3_corr_ratio_U0}d, with $\lambda_\mathrm{c} = 0.130 \pm 0.004$ for the 
\gls*{BOW} transition and no crossing for the \gls*{AFM} correlation ratio. The similarity of the results in Figs.~\ref{fig:figure3_corr_ratio_U0}b and \ref{fig:figure3_corr_ratio_U0}d suggest that the absence of \gls*{AFM} order is due to the 
nature of the \gls*{OSSH} coupling rather than the phonon energy. 

\begin{figure}[t]
    \centering
    \includegraphics[width=\columnwidth]{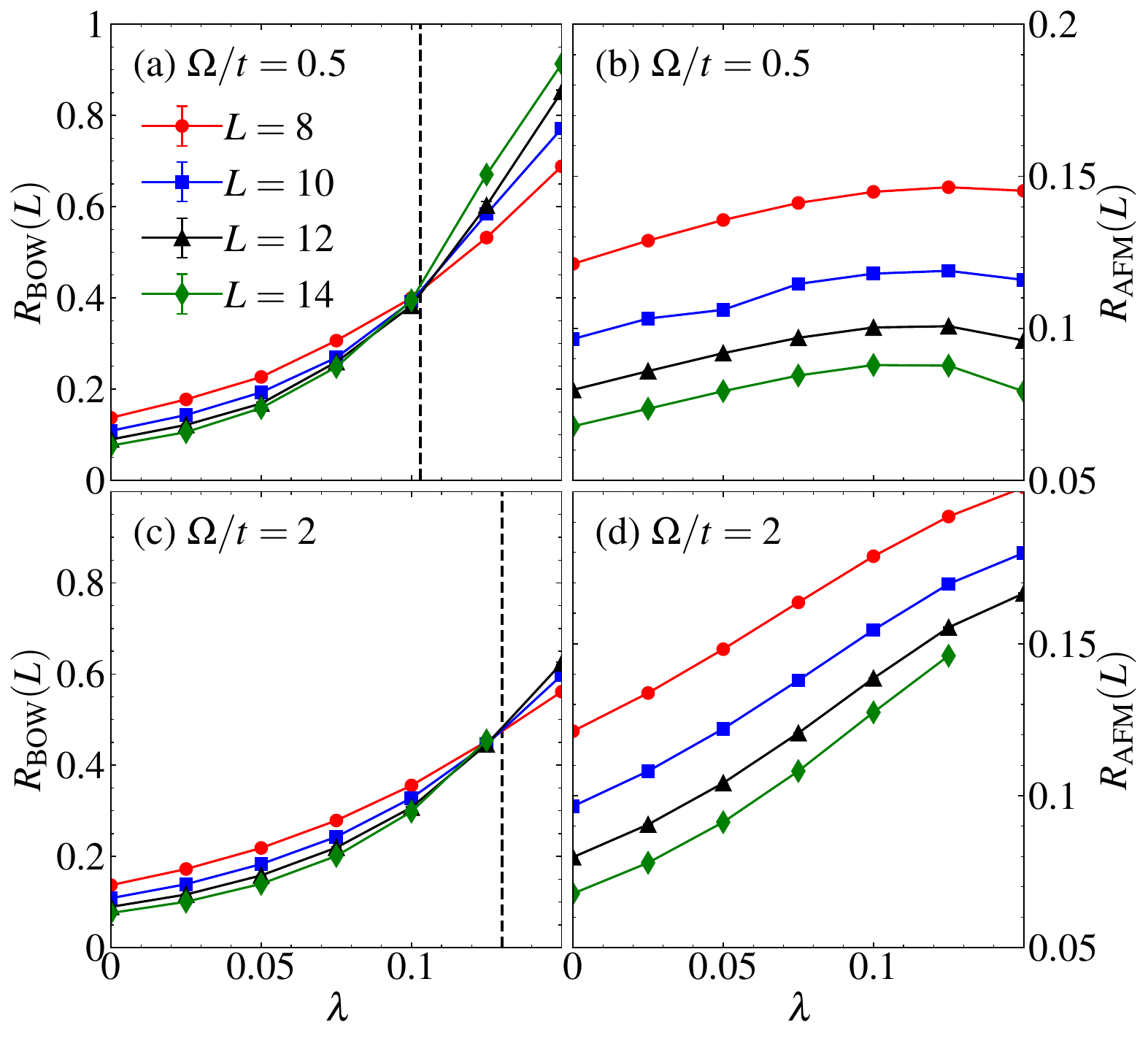}
    \caption{Correlation ratios for the bond-ordered-wave and antiferromagnetic orders as a function of $\lambda$ for the pure \gls*{OSSH} model ($U/t = 0$) at $\Omega/t = 0.5$ [panels (a) and (b)] and $\Omega/t = 2$ [panels (c) and (d)]. The \gls*{BOW} data indicates there is a finite critical coupling $\lambda_\mathrm{c}(\Omega/t = 0.5) = 0.103 \pm 0.010$ and $\lambda_\mathrm{c}(\Omega/t = 2) = 0.130 \pm 0.004$, while no crossing point is observed for the \gls*{AFM} correlations. All calculations were performed at an inverse temperature $\beta t = L$. 
    \label{fig:figure3_corr_ratio_U0}}
\end{figure}

The lack of \gls*{AFM} order in the \gls*{OSSH} model can be understood by considering the effective $e$-$e$ interaction that arises in the anti-adiabatic limit. Previous studies showed that in this limit, the \gls*{BSSH} model gives rise to a Heisenberg-like nearest-neighbor coupling between the electron spins that favors \gls*{AFM} correlations~\cite{Cai2021antiferromagnetism, Gotz2022Valence}. In Supplementary Note 1, we show that while the anti-adiabatic \gls*{OSSH} model retains this nearest-neighbor $e$-$e$ interaction, it also introduces longer-range interactions of similar strength between electrons on the same sublattice~\cite{supplement}. These longer-range interactions frustrate any \gls*{AFM} correlations from occurring. Intuitively, this difference in the effective $e$-$e$ interaction arises from the fact that each phonon mode in the \gls*{OSSH} model modulates two hopping integrals as opposed to just one, as in the \gls*{BSSH} model.

Our scaling analysis of the crossing ratio fixed $\beta = L$ to access low-temperature (ground state) properties. It is possible that our simulations are not reaching low enough temperatures to access the true ground state and that \gls*{AFM} does ultimately appear at lower temperatures. However, our results clearly show that if this is the case, the transition temperature is much lower in the \gls*{OSSH} model than in the \gls*{BSSH} model. This result is consistent with the presence of frustrating next-nearest neighbor interactions, as discussed in Supplementary Note 1.
 Ground state calculations would be valuable in this context. \\

\begin{figure}[t]
    \centering
    \includegraphics[width=\columnwidth]{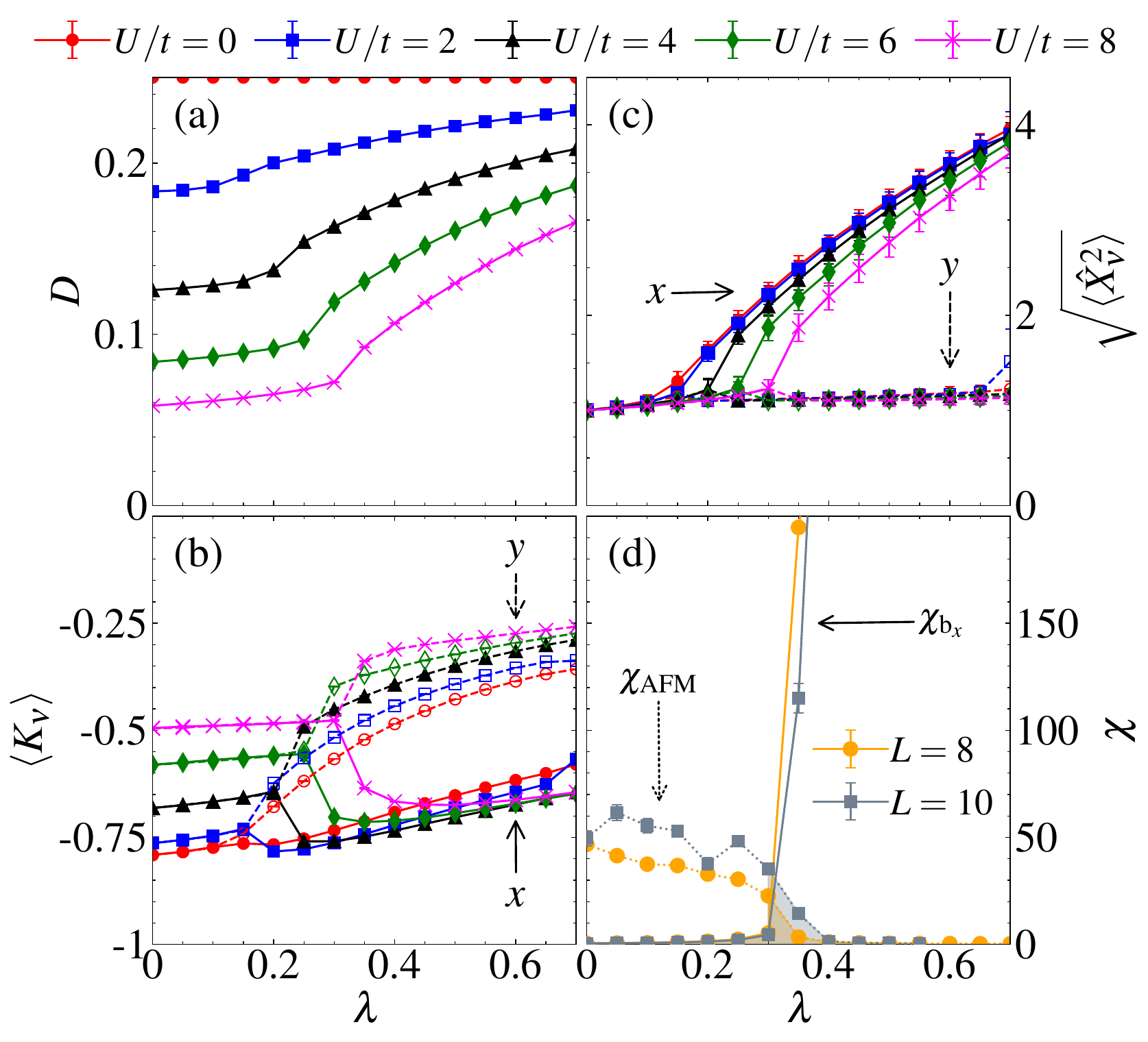}
    \caption{A summary of several local observables for the optical \gls*{SSH}-Hubbard model. (a) the average double occupancy $D$. (b) the average kinetic energy $\langle K_\nu\rangle$ in the $\hat{x}$ (solid lines) and $\hat{y}$ (dashed lines) directions. (c) the root mean square of the average displacement $(\langle \hat{X}_\nu^2\rangle)^{1/2}$ in the $\hat{x}$ and $\hat{y}$- directions.  (d) a comparison of the \gls*{AFM} (dotted lines) and $x$-direction \gls*{BOW} (solid lines) susceptibilities for fixed $U/t = 8$ at $L = 8$ (orange circles) and $L = 10$ (gray squares). The shaded area indicates the region of \gls*{AFM}/\gls*{BOW} coexistence. All results were obtained for $\Omega/t = 0.5$, $\beta t = 16$, and $\langle n\rangle = 1$. The results in panels a-c were obtained on $L = 8$ clusters. 
    \label{fig:figure3_observables}}
\end{figure}

To further examine the relevant transitions between \gls*{AFM} and \gls*{BOW} order and the possible coexistence region, we measured several local observables on $L=8$ clusters at low temperature ($\beta t = 16$). Local observables are often sensitive to changes in the dominant correlations in the system while being less prone to finite size effects~\cite{White1989, Johnston2013HubbardHolstein, Costa2020HHPhases}. 

Figure~\ref{fig:figure3_observables}a shows the average double occupancy $D = \langle n_{\boldsymbol{i},\uparrow} n_{\boldsymbol{i},\downarrow} \rangle$ as a function of $\lambda$ for selected values of $U/t$. Mott correlations for weak couplings initially suppress the double occupancy, but it grows as the strength of the \gls*{eph} coupling increases. Notably, the $D$ vs. $\lambda$ curves have a kink at $\lambda$ values that coincide with the transition from the \gls*{AFM} to \gls*{BOW} state, as indicated by the red line in Fig.~\ref{fig:figure1_phase_diagram}. For larger values of $\lambda$, the values of the double occupancies begin growing toward their uncorrelated values, which reflects the increased hybridization between pairs of sites in the \gls*{BOW} phase. 

The average values of the electron kinetic energy $\langle \hat{K}_\nu \rangle$ and the root mean square phonon displacement $(\langle \hat{X}^2_\nu\rangle)^{1/2}$ are shown in Figs.~\ref{fig:figure3_observables}b and \ref{fig:figure3_observables}c, respectively. 
Because the \gls*{SSH} interaction modulates the nearest-neighbor hopping integrals, both quantities indicate how strongly the lattice distorts across parameter space. 
We observe a bifurcation in both at coupling values that approximately coincide with \gls*{BOW} transition line shown in Fig.~\ref{fig:figure1_phase_diagram}. The average electron kinetic energy and root mean square phonon displacement measured along each crystallographic direction are equal in the \gls*{AFM} phase but split in the \gls*{BOW} phase. 
This bifurcation is also observed in \gls*{BSSH} models~\cite{Cai2021antiferromagnetism, TanjaroonLy2023comparative, Casebolt2024magnetic} and reflects the broken $C_4$ symmetry of the \gls*{BOW} phase, sketched in the right inset of Fig.~\ref{fig:figure1_phase_diagram}. 

Finally, Fig.~\ref{fig:figure3_observables}d examines the \gls*{AFM} and \gls*{BOW} susceptibilities for $U/t = 8$ and $L = 8$ and $10$ clusters. Here, we observe a small region where both susceptibilities have large non-zero values, as indicated by the shaded area. This behavior is consistent with coexistence between the \gls*{AFM} and \gls*{BOW} phases.\\

\begin{figure}[t]
    \centering
    \includegraphics[width=\columnwidth]{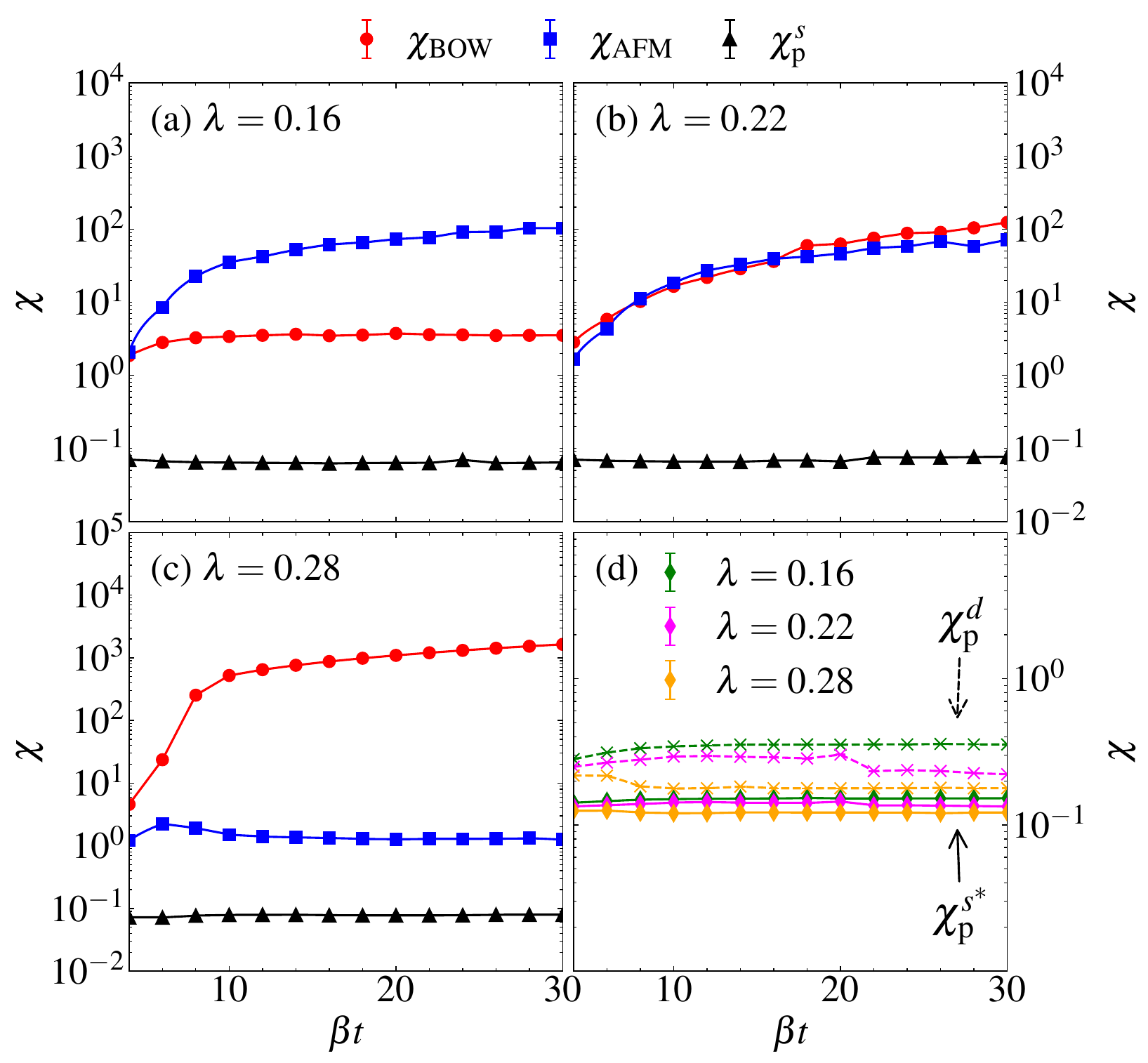}
    \caption{Finite temperature susceptibilities in different regions of the phase diagram. Panels (a), (b), and (c) show \gls*{AFM} (blue squares), \gls*{BOW} (red circles), and on-site $s$-wave pairing (black triangles) susceptibilities at $\lambda = 0.16$, $0.22$, and $0.28$, respectively. Panel (d) plots the corresponding extended $s$- ($\chi_\text{p}^{s*}$, solid lines) and $d$-wave ($\chi_\text{p}^{d}$, dashed lines) susceptibilities. Lines are a guide to the eye. All results were obtained from $L=14$ lattices with fixed $U/t = 4.5$, $\Omega = t/2$, and $\langle n\rangle = 1$.
    \label{fig:figure4_beta_sweep}}
\end{figure}

Next, we examine the temperature dependence of the \gls*{AFM}, \gls*{BOW}, and pairing susceptibilities $\chi_\gamma$ in different regions of the phase diagram. Fig.~\ref{fig:figure4_beta_sweep} plots the results on a log scale for $\lambda = 0.16, 0.22$, and $0.28$ as a function of $\beta t$ and fixed $U/t = 4.5$. These results were obtained on a larger $L = 14$ cluster. For $\lambda = 0.16$, we expect the system to be antiferromagnetic, which is corroborated by the dominant $\chi_\mathrm{s}(\pi,\pi)$ at low-temperatures shown in Fig.~\ref{fig:figure4_beta_sweep}a. Similarly, for $\lambda = 0.28$, we expect the system to be in a \gls*{BOW} phase, which is consistent with the dominant $\chi_\mathrm{b}(\pi,\pi)$ shown in Fig.~\ref{fig:figure4_beta_sweep}c. More interesting is the $\lambda = 0.22$ case, which lies within the coexistence region. In this case, the \gls*{AFM} and \gls*{BOW} susceptibilities are of the same order of magnitude and grow concomitantly with decreasing temperature. 

It has recently been shown that \gls*{BSSH}-like interactions can mediate strong binding between carriers in the dilute limit~\cite{Sous2018light} and potential high-temperature superconductivity, provided that it is not preempted by the formation of any competing order~\cite{Zhang2023, cai2023hightemperature}. The black triangles in Figs.~\ref{fig:figure4_beta_sweep}a-c plot the on-site $s$-wave pairing susceptibility as a function of temperature and give no indication of a superconducting instability. Fig~\ref{fig:figure4_beta_sweep}d also plots the temperature dependence of the extended $s$-wave and $d$-wave 
susceptibilities for the three regions identified in the phase diagram. Here we find no indications of robust $s$-wave, extended $s$-wave, or $d$-wave pairing. While the $d$-wave susceptibility is the dominant pairing channel for all values of $\lambda$ we have checked, its magnitude is several orders of magnitude smaller than either the \gls*{AFM} or \gls*{BOW} susceptibilities at low temperatures. 

The \gls*{AFM} phase found at small $U/t$ and $\lambda$ is metallic at $\beta = 16/t$. We have found that insulating behavior sets in upon further cooling, and we have observed no indications of low-temperature superconducting correlations in this region of the phase diagram either. These results provide strong numerical evidence that the \gls*{AFM} or \gls*{BOW} phases are dominant at half-filling across the entire ($\lambda, U$)-plane as might be expected for such a commensurate filling. It remains to be seen whether doping the system, which will introduce a sign problem, will induce superconductivity; however, we also note that a recent study of the pure doped \gls*{OSSH} model, which does not have a sign problem, found no indication of high-temperature superconductivity for these parameters~\cite{TanjaroonLy2023comparative}. \\

\section{Discussion}\label{sec::discussion}
\noindent We have conducted a detailed \gls*{DQMC} study of the magnetic and bond-ordered phases of the half-filled \gls*{OSSH}-Hubbard model in the $(\lambda, U)$-plane, focusing on the challenging adiabatic regime relevant to most materials. For weak \gls*{eph} coupling, the system is dominated by \gls*{AFM} Mott correlations similar to those observed in the single-band Hubbard model. Increasing the \gls*{eph} coupling strength above a non-zero $U$-dependent critical value $\lambda_c^\text{BOW}$ induces a \gls*{BOW} phase. 
Additionally, while there is no sign of enhanced superconducting correlations anywhere in the phase diagram, we did find evidence for a narrow region of coexistence between the \gls*{AFM} and \gls*{BOW} phases. Finally, in the $U=0$ limit, we did not detect any long-range 
\gls*{AFM} phase in the pure \gls*{OSSH} model. We did, however, observe a weak enhancement of \gls*{AFM} correlations in this regime, indicating that such correlations are short-ranged at the temperatures we accessed~\cite{supplement}. In this context, we note that any sign-problem-free-model is expect to have a low-temperature instability away from a Fermi-liquid ground state~\cite{Grossman2023Robust}. Thus we expect either superconductivity, antiferromagnetism, or non-Fermi-liquid behavior at zero temperature. However, which instability ultimately occurs currently cannot be assessed from our result.  

Previous \gls*{DQMC} studies of the \gls*{BSSH}-Hubbard model have also observed a transition between \gls*{AFM} and \gls*{BOW} phases~\cite{Cai2022robustness, Feng2022phase}. These studies focused on higher energy phonon modes ($\Omega = 3t/2$ in Ref.~\cite{Cai2022robustness} and $\Omega = t$ in Ref.~\cite{Feng2022phase}) and considered a model where each hopping integral is modulated independently, unlike in our model. Neither study reported any evidence for a region of coexistence between the two orders. We believe this difference stems from the nature of the microscopic coupling~\cite{TanjaroonLy2023comparative, MalkarugeCosta2023comparative}. For example, in the anti-adiabatic limit, the \gls*{BSSH} model can generate \gls*{AFM} order through an effective nearest-neighbor Heisenberg-like interaction~\cite{Cai2021antiferromagnetism}. In contrast, the \gls*{OSSH} model introduces additional longer-range interaction terms which frustrate the \gls*{AFM} order~\cite{supplement}. The combination of both an effective nearest-neighbor Heisenberg-like interaction and longer-range frustrating interactions would also explain the presence of enhanced short-range \gls*{AFM} correlations but the absence of long-range \gls*{AFM} order. The \gls*{BSSH} model also has a non-zero coupling to the $\boldsymbol{q} = 0$ modes, which is forbidden in the case of the \gls*{OSSH} model~\cite{MalkarugeCosta2023comparative}. The mobility of the \gls*{OSSH} polarons also appears to be much smaller than those of the \gls*{BSSH} model~\cite{TanjaroonLy2023comparative}. 

Understanding the interplay of \gls*{eph} and Coulomb interactions has implications for the study of many strongly correlated materials. While both the Hubbard-Holstein and the \gls*{BSSH}-Hubbard models have been explored numerically, the \gls*{OSSH} model has received comparatively less attention. With newer state-of-the-art numerical methods, we have successfully mapped out the principal phases of the half-filled \gls*{OSSH}-Hubbard model. In the future, it would be interesting to dope the system and study the differences in the superconducting correlations obtained from either the parent \gls*{AFM} or \gls*{BOW} phases.\\


\noindent{\bf Acknowledgments}: A.~T.~L. would like to thank L. Rademaker for useful discussions and hospitality. This work was primarily supported by the National Science Foundation under Grant No.~DMR-2401388. A.~T.~L. acknowledges additional support from the U.S. Department of Energy, Office of Science, Office of Workforce Development for Teachers and Scientists, Office of Science Graduate Student Research (SCGSR) program, during the writing phase of this project. The SCGSR program is administered by the Oak Ridge Institute for Science and Education (ORISE) for the DOE. ORISE is managed by ORAU under contract number DE-SC0014664. This research used resources of the Compute and Data Environment for Science (CADES) at the Oak Ridge National Laboratory, which is supported by the Office of Science of the U.S. Department of Energy under Contract No. DE-AC05-00OR22725. \\


\noindent{\bf Data Availability}: The data supporting this study will be made available through a public repository once it is published. Until then, the data will be made available on request. The code supporting this study is available at: \href{https://github.com/SmoQySuite/SmoQyDQMC.jl}{https://github.com/SmoQySuite/SmoQyDQMC.jl}. \\


\bibliography{references.bib}

\end{document}


\title{Supplementary materials for ``Antiferromagnetic and bond-order-wave phases in the half-filled two-dimensional optical Su-Schrieffer-Heeger-Hubbard model''}
\date{\today}

\author{Andy~Tanjaroon~Ly\orcidlink{0000-0003-3162-8581}}
\affiliation{Department of Physics and Astronomy, The University of Tennessee, Knoxville, TN 37996, USA}
\affiliation{Institute for Advanced Materials and Manufacturing, The University of Tennessee, Knoxville, TN 37996, USA\looseness=-1} 

\author{Benjamin~Cohen-Stead\orcidlink{0000-0002-7915-6280}}
\affiliation{Department of Physics and Astronomy, The University of Tennessee, Knoxville, TN 37996, USA}
\affiliation{Institute for Advanced Materials and Manufacturing, The University of Tennessee, Knoxville, TN 37996, USA\looseness=-1} 

\author{Steven Johnston\orcidlink{0000-0002-2343-0113}}
\affiliation{Department of Physics and Astronomy, The University of Tennessee, Knoxville, TN 37996, USA}
\affiliation{Institute for Advanced Materials and Manufacturing, The University of Tennessee, Knoxville, TN 37996, USA\looseness=-1} 

{
\let\clearpage\relax
\maketitle
}
\large
\noindent\textbf{Supplementary Note 1: Effective interaction in the anti-adiabatic limit\label{app:eff_ee}}\normalsize \\

The action of the \gls*{OSSH} model on the imaginary time axis is given by
\begin{equation*}
    S\left[\bar{\psi},\psi,X\right]=\int\mathrm{d}\tau\sum_{\boldsymbol{i},\nu}\left[\bar{\psi}\partial_{\tau}\psi-t\hat{B}_{\boldsymbol{i},\boldsymbol{i}+\boldsymbol{a_{\nu}}}+\frac{M}{2}\left(\partial_{\tau}\hat{X}_{\boldsymbol{i},\nu}\right)^{2}+\frac{K}{2}X_{\boldsymbol{i},\nu}^{2}+\alpha\left(\hat{X}_{\boldsymbol{i}+\boldsymbol{a}_{\nu},\nu}-\hat{X}_{\boldsymbol{i},\nu}\right)\hat{B}_{\boldsymbol{i},\boldsymbol{i}+\boldsymbol{a_{\nu}}}\right], 
\end{equation*}
where $\nu$ runs over the $\hat{x}$ and $\hat{y}$ spatial dimensions. Here, $\boldsymbol{a}_{x}=(a,0)$ and $\boldsymbol{a}_{y}=(0,a)$ are the primitive lattice vectors, $\hat{X}_{\boldsymbol{i},\nu}$ is the position operator for the atom at site $\boldsymbol{i}$ along each of the directions $\nu$ of the lattice, and $\hat{B}_{\boldsymbol{i},\boldsymbol{i}+\boldsymbol{a}_{\nu}}=\sum_{\sigma}\left(\bar{\psi}_{\boldsymbol{i}}\psi_{\boldsymbol{i+\boldsymbol{a}_{\nu}}}+\mathrm{H.c.}\right)$ is the bond operator between lattice sites $\boldsymbol{i}$ and $\boldsymbol{i}+\boldsymbol{a}_{\nu}$. 

In the anti-adiabatic limit, $M\to0$ (or $\Omega\to\infty$) and the phonon kinetic term drops out of the integral
\begin{equation*}
    S\left[\bar{\psi},\psi,X\right]=\int\mathrm{d}\tau\sum_{\boldsymbol{i},\nu}\left[\bar{\psi}\partial_{\tau}\psi-t\hat{B}_{\boldsymbol{i},\boldsymbol{i}+\boldsymbol{a_{\nu}}}+\frac{K}{2}X_{\boldsymbol{i},\nu}^{2}+\alpha\left(\hat{X}_{\boldsymbol{i}+\boldsymbol{a}_{\nu},\nu}-\hat{X}_{\boldsymbol{i},\nu}\right)\hat{B}_{\boldsymbol{i},\boldsymbol{i}+\boldsymbol{a_{\nu}}}\right].
\end{equation*}

Next, we redefine the \gls*{eph} interaction term to an equivalent form
\begin{equation*}
    \sum_{\boldsymbol{i},\nu} \alpha\left(\hat{X}_{\boldsymbol{i}+\boldsymbol{a}_{\nu},\nu}-\hat{X}_{\boldsymbol{i},\nu}\right)\hat{B}_{\boldsymbol{i},\boldsymbol{i}+\boldsymbol{a_{\nu}}}= \sum_{\boldsymbol{i},\nu} \alpha\hat{X}_{\boldsymbol{i},\nu}\left(\hat{B}_{\boldsymbol{i},\boldsymbol{i}+\boldsymbol{a_{\nu}}}-\hat{B}_{\boldsymbol{i},\boldsymbol{i}-\boldsymbol{a_{\nu}}}\right).
\end{equation*}
This form allows us to complete the square and obtain 
\begin{equation*}
    \begin{split}
      S\left[\bar{\psi},\psi,X\right]&=\int\mathrm{d}\tau\sum_{\boldsymbol{i},\nu}\left[\bar{\psi}\partial_{\tau}\psi-t\hat{B}_{\boldsymbol{i},\boldsymbol{i}+\boldsymbol{a_{\nu}}}+\frac{K}{2}X_{\boldsymbol{i},\nu}^{2}+\alpha\hat{X}_{\boldsymbol{i},\nu}\left(\hat{B}_{\boldsymbol{i},\boldsymbol{i}+\boldsymbol{a_{\nu}}}-\hat{B}_{\boldsymbol{i},\boldsymbol{i}-\boldsymbol{a_{\nu}}}\right)\right]\\&=\int\mathrm{d}\tau\sum_{\boldsymbol{i},\nu}\left\{ \bar{\psi}\partial_{\tau}\psi-t\hat{B}_{\boldsymbol{i},\boldsymbol{i}+\boldsymbol{a_{\nu}}}+\frac{K}{2}\left[X_{\boldsymbol{i},\nu}+\frac{\alpha}{K}\left(\hat{B}_{\boldsymbol{i},\boldsymbol{i}+\boldsymbol{a_{\nu}}}-\hat{B}_{\boldsymbol{i},\boldsymbol{i}-\boldsymbol{a_{\nu}}}\right)\right]^{2}-\frac{\alpha^{2}}{K}\left(\hat{B}_{\boldsymbol{i},\boldsymbol{i}+\boldsymbol{a_{\nu}}}-\hat{B}_{\boldsymbol{i},\boldsymbol{i}-\boldsymbol{a_{\nu}}}\right)^{2}\right\} \\&=\int\mathrm{d}\tau\sum_{\boldsymbol{i},\nu}\left[\bar{\psi}\partial_{\tau}\psi-t\hat{B}_{\boldsymbol{i},\boldsymbol{i}+\boldsymbol{a_{\nu}}}-\frac{\alpha^{2}}{K}\left(\hat{B}_{\boldsymbol{i},\boldsymbol{i}+\boldsymbol{a_{\nu}}}-\hat{B}_{\boldsymbol{i},\boldsymbol{i}-\boldsymbol{a_{\nu}}}\right)^{2}\right], 
    \end{split}
\end{equation*} 
where we have expliclty integrated out the phonons in the last step. 
Expanding the $\left(\hat{B}_{\boldsymbol{i},\boldsymbol{i}+\boldsymbol{a_{\nu}}}-\hat{B}_{\boldsymbol{i},\boldsymbol{i}-\boldsymbol{a_{\nu}}}\right)^{2}$ term, we then obtain
\begin{equation*}
    \begin{split}
        S\left[\bar{\psi},\psi,X\right]&=\int\mathrm{d}\tau\sum_{\boldsymbol{i},\nu}\left[\bar{\psi}\partial_{\tau}\psi-t\hat{B}_{\boldsymbol{i},\boldsymbol{i}+\boldsymbol{a_{\nu}}}-\frac{\alpha^{2}}{K}\left(\hat{B}_{\boldsymbol{i},\boldsymbol{i}+\boldsymbol{a_{\nu}}}^{2}+\hat{B}_{\boldsymbol{i},\boldsymbol{i}-\boldsymbol{a_{\nu}}}^{2}-\hat{B}_{\boldsymbol{i},\boldsymbol{i}+\boldsymbol{a_{\nu}}}\hat{B}_{\boldsymbol{i},\boldsymbol{i}-\boldsymbol{a_{\nu}}}-\hat{B}_{\boldsymbol{i},\boldsymbol{i}-\boldsymbol{a_{\nu}}}\hat{B}_{\boldsymbol{i},\boldsymbol{i}+\boldsymbol{a_{\nu}}}\right)\right]\\
        &=\int\mathrm{d}\tau\sum_{\boldsymbol{i},\nu}\left[\bar{\psi}\partial_{\tau}\psi+\left(-t\hat{B}_{\boldsymbol{i},\boldsymbol{i}+\boldsymbol{a_{\nu}}}-\frac{\alpha^{2}}{K}\hat{B}_{\boldsymbol{i},\boldsymbol{i}+\boldsymbol{a_{\nu}}}^{2}\right)-\frac{\alpha^{2}}{K}\left(\hat{B}_{\boldsymbol{i},\boldsymbol{i}-\boldsymbol{a_{\nu}}}^{2}-\hat{B}_{\boldsymbol{i},\boldsymbol{i}+\boldsymbol{a_{\nu}}}\hat{B}_{\boldsymbol{i},\boldsymbol{i}-\boldsymbol{a_{\nu}}}-\hat{B}_{\boldsymbol{i},\boldsymbol{i}-\boldsymbol{a_{\nu}}}\hat{B}_{\boldsymbol{i},\boldsymbol{i}+\boldsymbol{a_{\nu}}}\right)\right]\\
        &=\int\mathrm{d}\tau\sum_{\boldsymbol{i},\nu}\left[\bar{\psi}\partial_{\tau}\psi-\left(t\hat{B}_{\boldsymbol{i},\boldsymbol{i}+\boldsymbol{a_{\nu}}}-\frac{2\alpha^{2}}{K}\hat{B}_{\boldsymbol{i},\boldsymbol{i}+\boldsymbol{a_{\nu}}}^{2}\right)+\frac{\alpha^{2}}{K}\left(\hat{B}_{\boldsymbol{i},\boldsymbol{i}+\boldsymbol{a_{\nu}}}\hat{B}_{\boldsymbol{i},\boldsymbol{i}-\boldsymbol{a_{\nu}}}+\hat{B}_{\boldsymbol{i},\boldsymbol{i}-\boldsymbol{a_{\nu}}}\hat{B}_{\boldsymbol{i},\boldsymbol{i}+\boldsymbol{a_{\nu}}}\right)\right].
    \end{split}
\end{equation*}

The term $-t\hat{B}_{\boldsymbol{i},\boldsymbol{i}+\boldsymbol{a_{\nu}}}-\frac{2\alpha^{2}}{K}\hat{B}_{\boldsymbol{i},\boldsymbol{i}+\boldsymbol{a_{\nu}}}^{2}$ is also obtained in the \gls*{BSSH} model and it produces an effective Heisenberg-like Hamiltonian with an antiferromagnetic $J=2\alpha^2/K$~\cite{Cai2021antiferromagnetism}. Here, however, we have additional terms proportional to $\alpha^2/K$. We can expand the products of operators and rearrange terms 

\begin{align*}
    \frac{\alpha^{2}}{K}\sum_{\boldsymbol{i},\nu}\left(\hat{B}_{\boldsymbol{i},\boldsymbol{i}+\boldsymbol{a}_{\nu}}\hat{B}_{\boldsymbol{i},\boldsymbol{i}-\boldsymbol{a}_{\nu}}+\hat{B}_{\boldsymbol{i},\boldsymbol{i}-\boldsymbol{a}_{\nu}}\hat{B}_{\boldsymbol{i},\boldsymbol{i}+\boldsymbol{a}_{\nu}}\right)
    &=-\frac{2\alpha^{2}}{K}\left[\sum_{\boldsymbol{i},\nu}\sum_{\sigma}\left(c_{\boldsymbol{i-\boldsymbol{a}_{\nu}},\sigma}^{\dagger}n_{\boldsymbol{i},\sigma}c_{\boldsymbol{i+\boldsymbol{a}_{\nu},\sigma}}+c_{\boldsymbol{i+\boldsymbol{a}_{\nu}},\sigma}^{\dagger}n_{\boldsymbol{i},\sigma}c_{\boldsymbol{i-\boldsymbol{a}_{\nu},\sigma}}\right)\right.\\&\left.+\sum_{\boldsymbol{i},\nu}\sum_{\sigma\neq\sigma^{\prime}}\left(c_{\boldsymbol{i+\boldsymbol{a}_{\nu}},\sigma}^{\dagger}c_{\boldsymbol{i},\sigma^{\prime}}^{\dagger}c_{\boldsymbol{i},\sigma}c_{\boldsymbol{i-\boldsymbol{a}_{\nu},\sigma^{\prime}}}+c_{\boldsymbol{i-\boldsymbol{a}_{\nu}},\sigma}^{\dagger}c_{\boldsymbol{i},\sigma^{\prime}}^{\dagger}c_{\boldsymbol{i},\sigma}c_{\boldsymbol{i+\boldsymbol{a}_{\nu},\sigma^{\prime}}}\right)\right]\\&+\frac{\alpha^{2}}{K}\sum_{\boldsymbol{i},\nu}\hat{B}_{\boldsymbol{i+\boldsymbol{a}_{\nu}},\boldsymbol{i}-\boldsymbol{a}_{\nu}},
\end{align*}
where $\hat{B}_{\boldsymbol{i+\boldsymbol{a}_{\nu}},\boldsymbol{i}-\boldsymbol{a}_{\nu}}=\sum_{\sigma}\left(c_{\boldsymbol{i+\boldsymbol{a}_{\nu}},\sigma}^{\dagger}c_{\boldsymbol{i-\boldsymbol{a}_{\nu},\sigma}}+\mathrm{h.c}\right)=\sum_{\sigma}\left(\bar{\psi}_{\boldsymbol{i+\boldsymbol{a}_{\nu}},\sigma}\psi_{\boldsymbol{i-\boldsymbol{a}_{\nu},\sigma}}+\mathrm{h.c}\right)$. The first term represents a hopping interaction that allows an electron to hop between next-nearest neighbor sites along the $\hat{x}$ or $\hat{y}$ directions with amplitude $2\alpha^2/K$ depending upon the density of the intermediate site. The second term represents an interaction that shifts a bond pair by one site. The final term allows electrons to hop between next-nearest neighbors along the crystallographic directions with amplitude $\alpha^2/K$.

\clearpage
\large
\noindent\textbf{Supplementary Note 2: Compressibility and Spectral Weight}\normalsize \\

\begin{figure}[h]
    \centering
    \includegraphics[width=0.8 \columnwidth]{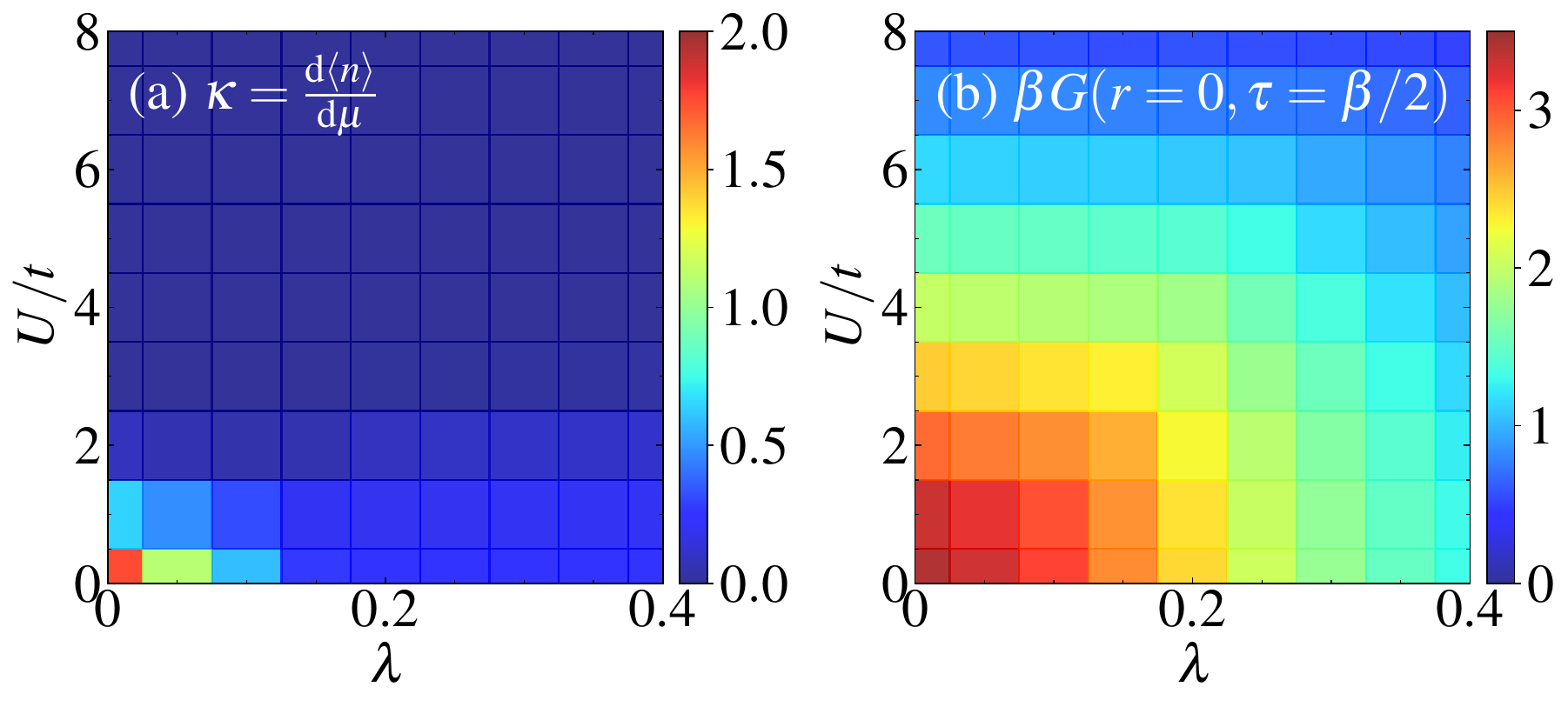}
    \caption{The (a) compressibility $\kappa$ and (b) local single particle  Green's function $\beta G(\boldsymbol{r}=0,\tau=\beta/2)$ for the optical SSH-Hubbard model as a function of $\lambda$ and $U$. All results are for an $L=8$ cluster at $\beta t = 16$ with $\omega/t = 0.5$.
    \label{fig:figureS1_compressibility_spectral}}
\end{figure}

To assess the transport properties of the system, we measured the system's compressibility $\kappa = \frac{\mathrm{d}\langle n\rangle}{\mathrm{d}\mu}$ and local Green's function $\beta G(\boldsymbol{r} = 0,\tau = \beta/2)$ for an $L=8$ cluster at $\beta t = 16$. The compressibility $\kappa = 0$ ($\ne 0$) for an insulating (metallic) system. Conversely, the local Green's function provides a measure of the integrated spectral weight near the Fermi energy~\cite{Trivedi1995Ddviations}. Fig.~\ref{fig:figureS1_compressibility_spectral}a plots the compressibility in the ($\lambda$, $U$)-plane. It is non-zero for $U\lesssim 2$ and $\lambda \lesssim 0.1$ indicating that this region of the phase diagram is metallic at this temperature. $\kappa = 0$ throughout the remainder of the phase diagram. Fig.~\ref{fig:figureS1_compressibility_spectral}b plots the single particle Green's function. It is also largest in the same region where $\kappa \ne 0$. \\

\clearpage
\large
\noindent\textbf{Supplementary Note 3: Correlation Ratios}\normalsize \\

\begin{figure}[h]
    \centering
    \includegraphics[width=\columnwidth]{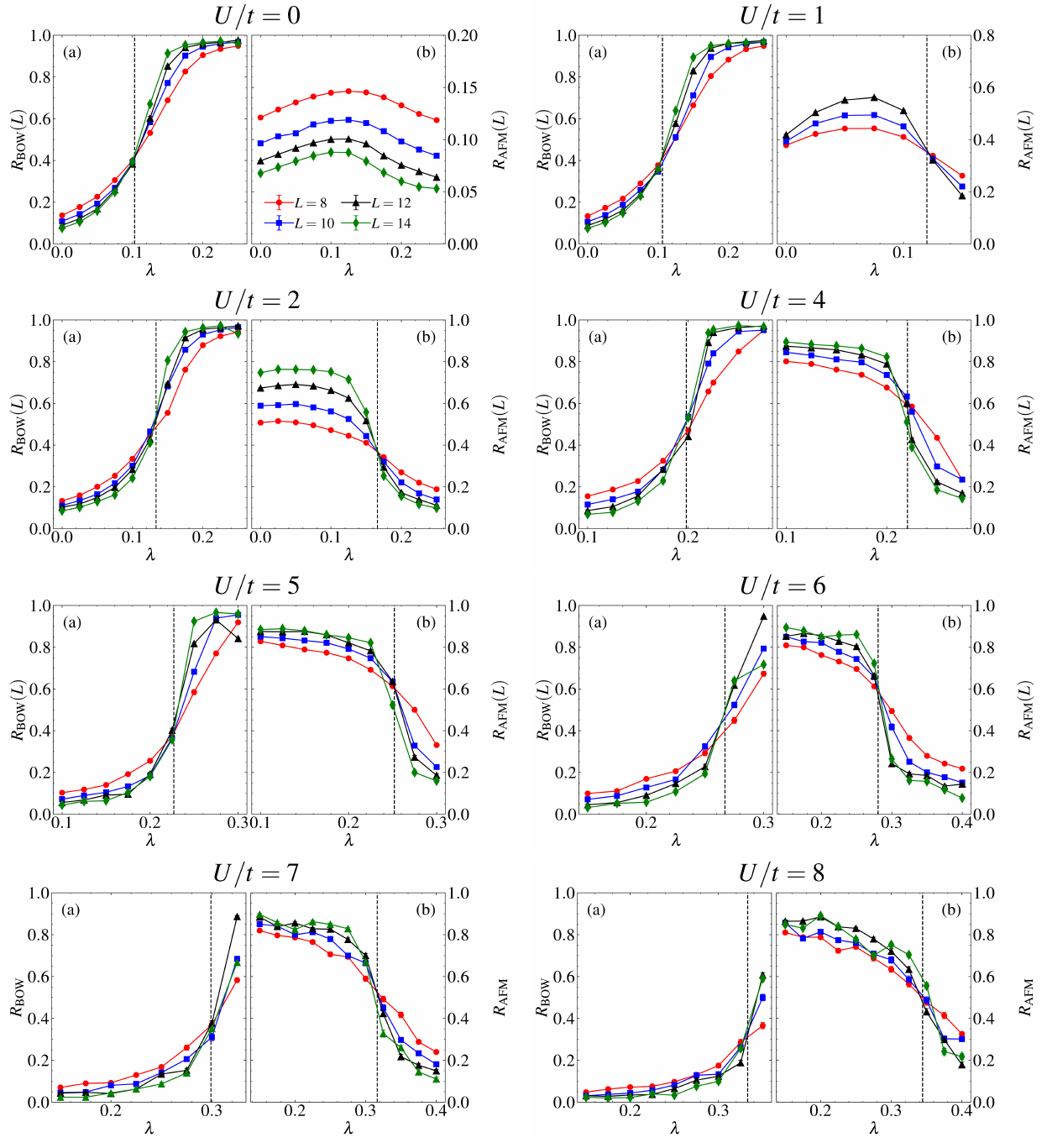}
    \caption{(a) Bond-ordered-wave and (b) antiferromagnetic (AFM) correlation ratios for $U/t\in[0,8]$, as indicated in each subplot. The dashed black line indicates the location of the quantum critical point (QCP).
    \label{fig:figureS2_corr_ratios}}
\end{figure}

Figure~\ref{fig:figureS2_corr_ratios} shows additional correlation ratio crossings for $U/t \in [0,8]$ as indicated in each panel. Each (a) panel plots the \gls*{BOW} correlation ratios while each (b) panel plots the \gls*{AFM} correlation ratios. 

\clearpage
\large
\noindent\textbf{Supplementary Note 4: Antiferromagnetic susceptibility and structure factor }\normalsize \\

\begin{figure}[h]
    \centering
    \includegraphics[width=0.6\columnwidth]{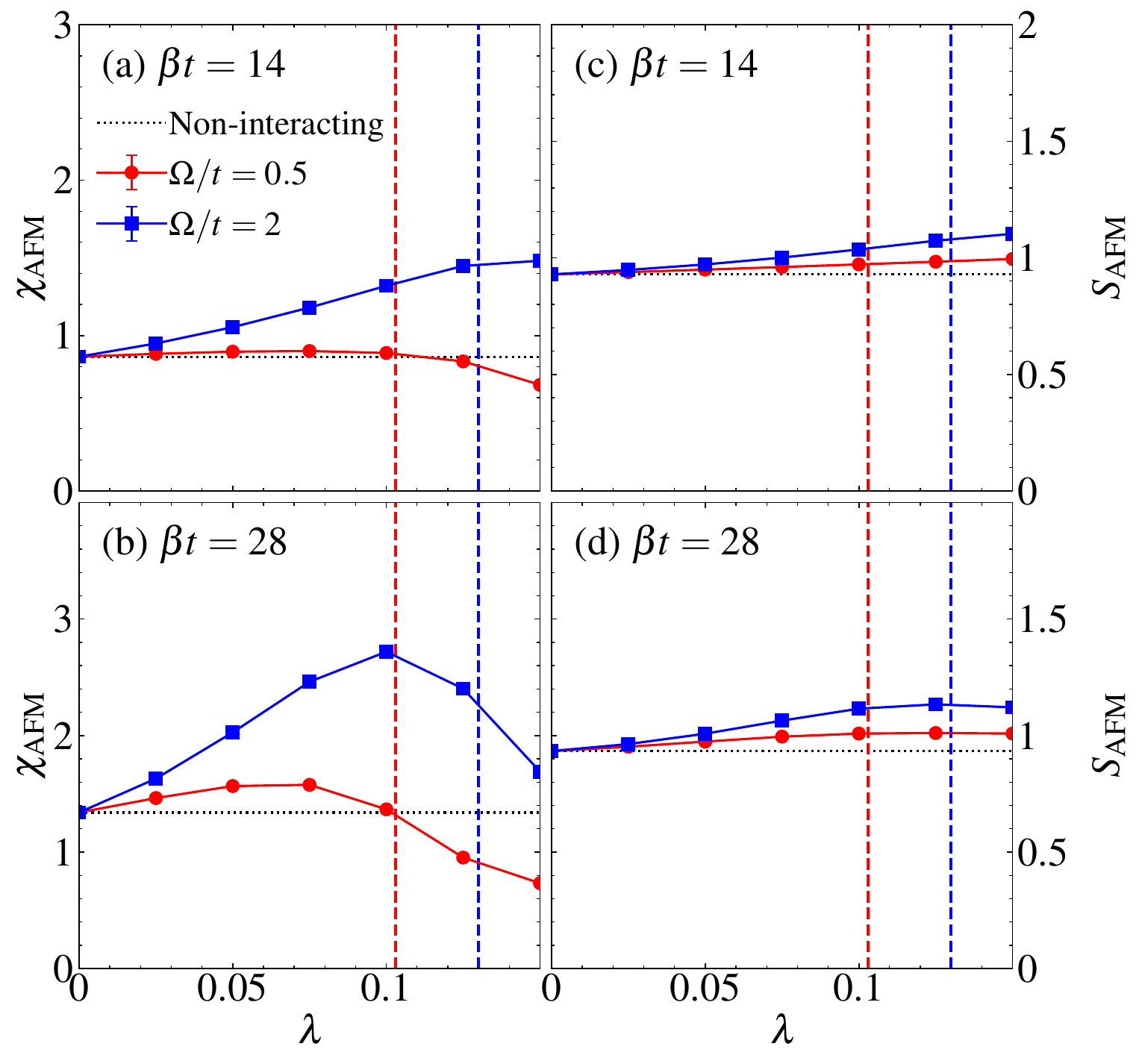}
    \caption{Antiferromagnetic susceptibility $\chi_{\mathrm{AFM}}$ for $L=14$, $\Omega/t=0.5 \text{ and } 2$, $\beta t = 14$ [panel (a)], and $\beta t = 28$ [panel (b)] compared to the non-interacting solution (black dotted line). The vertical dashed lines denote the value of $\lambda_{c}^{\mathrm{BOW}}$ for each respective phonon frequency. The corresponding antiferromagnetic structure factors $S_{\mathrm{AFM}}$ are shown in panels (c) and (d).
    \label{fig:figureS3_afm_susc_struc}}
\end{figure}

Here we assess the low temperature \gls*{AFM} susceptibility and equal-time structure factors of the pure \gls*{OSSH} model ($U/t = 0$). Figs.~\ref{fig:figureS3_afm_susc_struc}a and ~\ref{fig:figureS3_afm_susc_struc}b plot the \gls*{AFM} susceptibility $\chi_{\mathrm{AFM}}$ obtained on an $L = 14$ cluster for $\beta t = 14$ and $\beta t = 28$, respectively, and two different phonon energies. For $\Omega/t = 0.5$, there is a weak enhancement in $\chi_{\mathrm{AFM}}$ compared to the non-interacting ($\lambda = 0$) case. Increasing the phonon energy to $\Omega/t =2$ results in a slightly larger enhancement. 
The dashed lines in each panel indicate the bond-order-wave quantum critical points $\lambda_c^{\mathrm{BOW}}$ for the two phonon frequencies. In this case, $\chi_{\mathrm{AFM}}$ is suppressed with the onset of long-range \gls*{BOW} order.  

Figs.~\ref{fig:figureS3_afm_susc_struc}c and ~\ref{fig:figureS3_afm_susc_struc}d plot the corresponding structure factor $S_{\mathrm{AFM}}$. Here we also observe a slight increase in the strength of the \gls*{AFM} correlations but the degree of the enhancement is much smaller than in the susceptibility. Since the latter integrates over all imaginary time fluctuations, this behavior is consistent with the presence of increased short-range fluctuating \gls*{AFM} correlations and a lack of long-range magnetic order at this temperature. 

\clearpage
\bibliography{references.bib}